\documentclass[sigconf]{acmart}

\usepackage[utf8]{inputenc} % allow utf-8 input
\usepackage[T1]{fontenc}    % use 8-bit T1 fonts
\usepackage{hyperref}       % hyperlinks
\usepackage{url}            % simple URL typesetting
\usepackage{booktabs}       % professional-quality tables
\usepackage{amsfonts}       % blackboard math symbols
\usepackage{nicefrac}       % compact symbols for 1/2, etc.
\usepackage{microtype}      % microtypography
\usepackage{xcolor}         % colors

\usepackage{multirow}
\usepackage{subfigure}

\usepackage[english]{babel}
\usepackage{moresize}
\usepackage{amsmath}
\usepackage{algorithmic}
\usepackage{balance}
\usepackage{comment}
\usepackage{paralist}
\usepackage{bm}
\usepackage{pgfplots}
\usetikzlibrary{pgfplots.dateplot}

\usepackage{flushend}
\usepackage[english]{babel}
\usepackage{graphicx}

\usepackage{amssymb}
\usepackage{amsfonts}
\usepackage{url}
\usepackage{bbm}
\usepackage{longtable}
\usepackage{rotating}
\usepackage{multirow}
\usepackage{mathrsfs}
\usepackage{enumitem}
\usepackage[linesnumbered,algoruled,boxed,lined]{algorithm2e}
\usepackage{adjustbox}
\usepackage{hyperref}
\usepackage{pgfplots}
\usetikzlibrary{pgfplots.dateplot}
\usepackage{filecontents}

\newcommand{\ie}{\textit{i}.\textit{e}.}
\newcommand{\eg}{\textit{e}.\textit{g}.}

\def\model{RecDiff}

\newcommand{\hide}[1]{} %hide

\newcommand{\mata}{\textbf{A}}
\newcommand{\matd}{\textbf{D}}
\newcommand{\mate}{\textbf{E}}
\newcommand{\mati}{\textbf{I}}
\newcommand{\matn}{\textbf{N}}
\newcommand{\matr}{\textbf{R}}
\newcommand{\mats}{\textbf{S}}
\newcommand{\matw}{\textbf{W}}
\newcommand{\vece}{\textbf{e}}
\newcommand{\vech}{\textbf{h}}
\newcommand{\vecx}{\textbf{x}}
\newcommand{\vecb}{\textbf{b}}
\newcommand{\dmnr}{\mathbb{R}}
\newcommand{\setu}{\mathcal{U}}
\newcommand{\setv}{\mathcal{V}}
\newcommand{\sete}{\mathcal{E}}
\newcommand{\graph}{\mathcal{G}}
\newcommand{\param}{\mathbf{\Theta}}
\newcommand{\loss}{\mathcal{L}}
\newcommand{\gauss}{\mathcal{N}}
\newcommand{\calo}{\mathcal{O}}

\AtBeginDocument{%
  }

\setcopyright{none}
\begin{document}
\fancyhead{}
\title{RecDiff: Diffusion Model for Social Recommendation}
% \author{Anonymous Author(s)}

% \renewcommand{\shortauthors}{Anonymous Author(s)}

\author{Zongwei Li}
\affiliation{%
  \institution{University of Hong Kong}
  \city{Hong Kong}
  \country{China}
}
\email{zongwei9888@gmail.com}

\author{Lianghao Xia}
\affiliation{%
  \institution{University of Hong Kong}
  \city{Hong Kong}
  \country{China}}
\email{aka\_xia@foxmail.com}

\author{Chao Huang}
\authornote{Chao Huang is the Corresponding Author.}
\affiliation{%
  \institution{University of Hong Kong}
  \city{Hong Kong}
  \country{China}}
\email{chaohuang75@gmail.com}

\begin{abstract}
Social recommendation has emerged as a powerful approach to enhance personalized recommendations by leveraging the social connections among users, such as following and friend relations observed in online social platforms. The fundamental assumption of social recommendation is that socially-connected users exhibit homophily in their preference patterns. This means that users connected by social ties tend to have similar tastes in user-item activities, such as rating and purchasing. However, this assumption is not always valid due to the presence of irrelevant and false social ties, which can contaminate user embeddings and adversely affect recommendation accuracy. To address this challenge, we propose a novel diffusion-based social denoising framework for recommendation (\model). Our approach utilizes a simple yet effective hidden-space diffusion paradigm to alleivate the noisy effect in the compressed and dense representation space. By performing multi-step noise diffusion and removal, \model\ possesses a robust ability to identify and eliminate noise from the encoded user representations, even when the noise levels vary. The diffusion module is optimized in a downstream task-aware manner, thereby maximizing its ability to enhance the recommendation process. We conducted extensive experiments to evaluate the efficacy of our framework, and the results demonstrate its superiority in terms of recommendation accuracy, training efficiency, and denoising effectiveness. The source code for the model implementation is publicly available at: \textcolor{blue}{\url{https://github.com/HKUDS/RecDiff}}.
\end{abstract}

% \keywords{datasets, neural networks, gaze detection, text tagging}


\maketitle

\section{Introduction}
\label{sec:intro}

Personalized recommender systems are pivotal in inferring users' preferences for items and are extensively utilized in various online services, including e-commerce systems (\eg, Alibaba and Amazon) \cite{zhang2019deep, yeh2023planning} and content sharing platforms (\eg, Tiktok and Netflix)~\cite{liu2021concept}. To overcome the challenge posed by limited user-item interactions, recent research has incorporated social relation data among users as an additional source of information. Known as social recommendation, these approaches aim to uncover shared preference patterns among socially-connected users, thereby enhancing user representations and improving recommendation performance \cite{du2022socially}.

Various social recommendation models have emerged, utilizing different technologies such as matrix factorization \cite{yang2016social}, attention mechanism \cite{chen2019social}, and graph neural network \cite{wu2019neural}. These models encode user-item interactions and user-wise social connections in a latent space, collaboratively fusing representations to enhance preference modeling. Recent studies, including SAMN \cite{chen2019social} and EATNN \cite{chen2019efficient}, incorporate attention mechanisms for capturing user relationships through weighted aggregation. Additionally, graph neural networks have also gained attention, with proposed encoders like DiffNet \cite{wu2019neural}, DGRec \cite{song2019session}, GraphRec \cite{fan2019graph}, and DANSER \cite{wu2019dual} emphasizing the extraction of high-order social connectivities.

Despite significant progress in social recommendation, a persistent challenge is the presence of inherent noise in social information. While auxiliary social data captures user relationships reflecting shared interests, it can also include irrelevant or false social connections that contradict users' similarity in preferences. Real-world social networks often have social links such as acquaintances and colleagues, which are unrelated to shared interests, resulting in socially-connected users exhibiting diverse preference patterns. This noisy social information can misguide recommenders by overemphasizing similarities between socially-connected users with limited shared interests. State-of-the-art GNN-based methods are particularly susceptible to this misleading effect due to their information propagation process on noisy social edges.

While there are limited works that attempt to address denoising social information for recommendation, handcrafted self-supervised learning techniques may offer limited help in reducing ambiguity in social links. For instance, MHCN \cite{yu2021self} and KCGN \cite{huang2021knowledge} utilize local-global mutual information maximization to constrain the extracted semantics of specific social edges using social community features at the subgraph level. SDCRec \cite{du2022socially} and DcRec \cite{wu2022disentangled} employ self-discrimination tasks, leveraging the contrastive learning paradigm for social recommenders. However, their pre-defined self-supervised learning (SSL) tasks may not effectively align with the objective of denoising social recommendation in two aspects. \\\vspace{-0.12in}

\noindent (i) These approaches heavily depend on handcrafted targets that may contain noise. In particular, the local-global infomax paradigm assumes less noise in global information, but the global social context can undergo semantic shifts. Contrastive learning introduces noise through random permutations. \\\vspace{-0.12in}

\noindent (ii) These self-supervised learning tasks fail to consider the wide array of social noise types and intensities. Infomax and contrastive approaches solely concentrate on specific deviations in social-aware user preferences without distinguishing between varying levels of noisy social dependencies. For instance, users with different social behaviors may exhibit varying degrees of irrelevant or false social links when it comes to modeling similar interests.\\\vspace{-0.12in}

Based on the above discussions, two fundamental questions arise regarding our social information denoiser for recommendation:
\begin{itemize}[leftmargin=*]
\item How can we design suitable training targets that directly align with the denoising task for social recommendation? \\\vspace{-0.12in}
\item How can we effectively capture social dependencies in user preferences while accounting for varying levels of noisy connections?
\end{itemize}

To address the aforementioned challenges, we present a social diffusion model, called \model, which combines the distinguishing capabilities of generative diffusion models with effective denoising-based training objectives. Our \model\ model excels in capturing data distributions and filtering noise in social graph structures, enabling accurate modeling of user similarities based on their preferences. Additionally, we employ a refined noise diffusion and removal process, allowing our social recommender to effectively handle various types of connection noise.

Our \model\ addresses the challenge of eliminating noisy social information by leveraging a diffusion model within a joint modeling framework that encompasses social relationships among users and interaction between users and items. To begin, \model\ encodes the structural features from both the social and interaction graphs into low-dimensional embeddings. These embeddings serve as the foundation for subsequent refinement through our diffusion-based denoiser. The denoiser is trained using a multi-step noise diffusion and elimination process within the representation space. By sampling different diffusion steps, our approach gains exposure to a wide range of noise scales, enhancing its ability to handle various types of social noise. Compared to diffusion models operating directly on the original graph domain, our hidden-space diffusion paradigm exhibits higher efficiency and a more compact solution space. Finally, the revised social embeddings are merged with user-item interaction modeling, resulting in improved recommendations that incorporate denoised social information.

The key contributions of this paper are summarized as follows:
\begin{itemize}[leftmargin=*]

\item We present the \model\ framework, a novel approach that enhances social recommender systems by effectively denoising social connections among users with a diffusion model.

\item Within our \model\ framework, we introduce an effective and efficient hidden-space diffusion paradigm. Through multi-step noise propagation and removal training, our model acquires robust denoising capabilities, enabling it to effectively handle diverse social connections among users. As a result, our model produces accurate user preference representations.

\item We conduct a comprehensive empirical study of our \model\ framework, demonstrating its superiority in terms of recommendation performance and denoising capability. 
% To ensure reproducibility, we open-source our model implementation at the following link: \url{https://anonymous.4open.science/r/RecDiff-4B25/}

\end{itemize}

\section{Preliminaries}
\label{sec:model}

\subsection{Social-aware Collaborative Graph}
We denote the users as $\mathcal{U}=\{u\}$ and the items as $\mathcal{V}=\{v\}$. The user-item interactions are represented by $\matr \in \dmnr^{|\mathcal{U}|\times|\mathcal{V}|}$, while the user-user social relations are represented by $\mats\in\dmnr^{|\mathcal{U}|\times |\mathcal{U}|}$. The user-item interactions capture historical user behaviors on items, \eg, clicks and views. Each element $r_{u,v}\in\matr$ is set as 1 if an interaction is observed between user $u$ and item $v$, and $r_{u,v}=0$ otherwise. Similarly, the social relation between a pair of users $(u,u')$ is represented by the binary element $s_{u,u'}\in\mats$. Here, $s_{u,u'}=1$ denotes social connections like friendship or following, while $s_{u,u'}=0$ indicates no such observation between user $u$ and $u'$.

To capture high-order connections in user-item interactions and user-user social relations, we transform them into graph-structured forms. The collaborative graph for user-item interactions, denoted as $\graph_r=(\setu, \setv, \sete_r)$, represents the relationships between users and items, where $\sete_r={(u,v)|r_{u,v}=1}$ represents the edge set. Similarly, the social graph based on user connections, denoted as $\graph_s=(\setu, \sete_s)$, captures the social relations between users, with $\sete_s={(u,u')|s_{u,u'}=1}$ representing the edge set.

\subsection{Graph-based Social Recommender}
State-of-the-art social recommendation methods typically utilize GNN-based encoding functions on collaborative graphs and social graphs to learn representations for users and items. These representations capture preference patterns and enable accurate prediction of user-item interactions. The paradigm can be formulated as:
\begin{align}
    \label{eq:social_rec}
    \hat{r}_{u,v} = \textbf{Pred}(\vece_u, \vece_v),~
    \mate=\textbf{Agg}(\mate^r, \mate^s),~\mate^{*}=\textbf{Enc}(\graph_{*})
\end{align}
where $\textbf{E}^*$ denotes $\textbf{E}^r$ or $\textbf{E}^s$ and $\graph_*$ denotes $\graph_r$ or $\graph_s$ for simplicity.
The collaborative graph $\graph_r$ and the social graph $\graph_s$ are firstly encoded by the GNN-based encoding function $\textbf{Enc}(\cdot)$ to produce node representations $\mate^{r}\in\dmnr^{(|\setu|+|\setv|)\times d}$ and $\mate^s\in\dmnr^{|\setu|\times d}$, respectively. The dual-view embeddings are then aggregated by a pooling method $\textbf{Agg}(\cdot)$ to yield the final node embeddings $\mate\in\dmnr^{(|\setu|+|\setv|)\times d}$. The embeddings are finally used by a predicting function $\textbf{Pred}(\cdot)$, such as the dot-product operation, to output predictions $\hat{r}_{u,v}$ for user-item interactions.
\section{Methodology}
\label{sec:solution}
\begin{figure*}[t]
    \centering
    \includegraphics[width=0.97\textwidth]{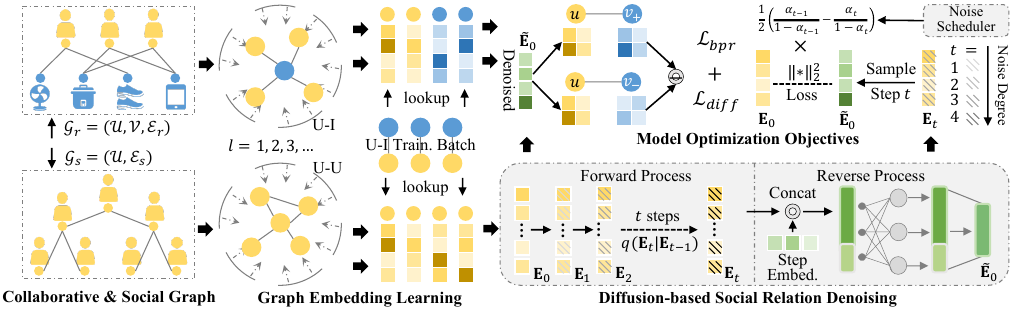}
    \vspace{-0.1in}
    \caption{Overall architecture of the proposed \model\ framework.}
    \vspace{-0.1in}
    \label{fig:framwork}
\end{figure*}

\subsection{Graph-based Collaborative Pattern Encoder}
Taking inspiration from the success of simplified Graph Neural Networks (GNNs)~\cite{he2020lightgcn}, we utilize a lightweight Graph Convolutional Network (GCN) as the graph encoder in our social denoising framework. This encoder operates on the user-item interaction graph $\graph_r$ and follows an iterative message passing process described as:
\begin{align}
    \label{eq:rec_enc}
    \mate^r = \sum_{l=0}^L \mate^r_l, ~~\mate^r_l = {h}(\matd^{-1/2}\mata^r\matd^{-1/2}\mate^r_{l-1}\matw)
\end{align}
Here, $\mata^r\in\dmnr^{(|\setu|+|\setv|)\times(|\setu|+|\setv|)}$ denotes the adjacency matrix of the collaborative graph $\graph_r$, and $\matd$ is the corresponding diagonal degree matrix. The embedding matrix $\mate^r_l\in\dmnr^{(|\setu|+|\setv|)\times d}$ captures the embeddings in the $l$-th iteration of the GCN. The initial embeddings, $\mate^r_0$, are randomly generated learnable parameters. In each iteration, the current embeddings are propagated to their neighboring nodes using the Laplacian-normalized adjacency matrix, and the $l_2$ embedding normalization function $h(\cdot)$ is applied. After $L$ iterations, the final user and item embeddings, $\mate^r$, are obtained by element-wise summation of the multi-order node embeddings.

\subsection{Diffusion-based Social Relation Denoising}
\subsubsection{\bf Hidden-Space Social Diffusion}
\begin{figure}[t]
    \centering
    \includegraphics[width=\columnwidth]{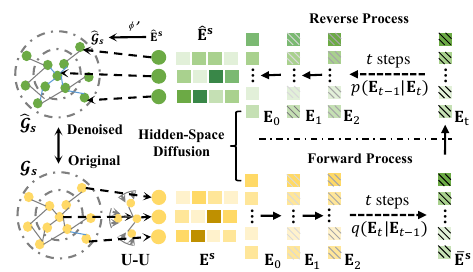}
    \caption{Illustration for the hidden-space social diffusion.}
    \vspace{-0.2in}
    \label{fig:hid_diff}
\end{figure}

Drawing inspiration from the success of diffusion models in generating noise-free data across various domains, such as images~\cite{saharia2022palette} and text~\cite{li2022diffusion}, our \model\ framework introduces a diffusion model to generate denoised social relation data. Considering the inherent sparsity of social graph data, we propose an approach that enables efficient and effective social diffusion by conducting forward and reverse diffusion processes in the latent space, rather than the graph data space. Figure~\ref{fig:hid_diff} provides an illustration of this hidden-space social diffusion mechanism, and its process can be summarized by the following formula:
\begin{align}
    \label{eq:hidden_project}
    \graph_s \stackrel{\phi}{\longrightarrow} \mate^s \stackrel{\psi}{\longrightarrow} \tilde{\mate}^s \stackrel{\psi'}{\longrightarrow} \hat{\mate}^s \stackrel{\phi'}{\longrightarrow} \hat{\graph}_s
\end{align}
Here, $\phi$ and $\phi'$ represent bidirectional projections between the graph data domain and the hidden embedding domain. $\psi$ and $\psi'$ refer to the forward and reverse processes of the diffusion model, where $\psi$ introduces noise to $\mate^s$, and $\psi'$ is optimized to remove this noise in the hidden space.

In graph-based recommendation, the encoding function $\textbf{Enc}(\cdot)$ and the predictive function $\textbf{Pred}(\cdot)$ are crucial for accurately compressing and reconstructing graph structures. By regarding them as the aforementioned projections $\phi$ and $\phi'$, respectively, these paired projections become invertible to each other. This property implies that by learning the hidden-space denoiser $\psi'$, our hidden-space diffusion model can effectively filter out noise present in the graph-structured data $\mathcal{G}$. Furthermore, due to the low-dimensional nature of the hidden space, training the hidden-space denoiser $\psi'$ is significantly easier compared to directly denoising in the graph data space. Building on the aforementioned considerations, we utilize a similar GCN model as the learnable $\phi$ projection, following the formulation in Eq~\ref{eq:rec_enc}. Subsequently, we design our social denoiser based on the acquired embeddings $\mate^s\in\dmnr^{|\setu|\times d}$.

\subsubsection{\bf Forward and Reverse Diffusion}
Based on the social relation data $\mate^s$ in the latent embedding space, we design the forward and reverse processes for our hidden-space diffusion. In the forward process, Gaussian noise is incrementally added to the original data $\mate^s$ until it eventually transforms into complete Gaussian noise. Conversely, the backward process utilizes trainable neural networks to eliminate noise, enabling the generation of noise-less social data.\\\vspace{-0.12in}

\noindent\textbf{Forward Process}. In the forward process, our \model\ iteratively applies $T$ noise steps, where $T$ is a hyperparameter. The data at the $t$-th step is denoted as $\mate_t$ (omitting the superscript $s$ for simplicity), and the 0-step data is the original data, \ie, $\mate_0=\mate^s$. The $t$-step data is calculated from the $(t-1)$-step data as follows:
\begin{align}
    \label{eq:gauss_step}
    q(\mate_t|\mate_{t-1}) = \gauss(\mate_t; \sqrt{1-\beta_t} \mate_{t-1}, \beta_t \mati)
\end{align}
Here, $\gauss$ represents the Gaussian distribution. The parameter $\beta_t$ controls the magnitude of the noise. It has been shown that by gradually increasing $\beta_t$ for $t=1,\cdots,T$, the noised data $\mate_t$ converges to complete Gaussian noise as $t$ increases~\cite{ho2020denoising}. This property allows our noise diffusion process to cover a wide range of noise levels in the data. To generate $\beta_t$, we use a linear interpolation sequence $s$ between two hyperparameters $\bar{s}{max}$ and $\bar{s}{min}$:
\begin{align}
    \label{eq:beta_gen}
    \beta_t = 1 - s_t / s_{t-1},~~ s=(1,\bar{s}_{max},\cdots, \bar{s}_{min})%,~~~\bar{s}_{max}>\bar{s}_{min}
\end{align}
Due to the additivity property of Gaussian distributions, the $t$-step data can be directly calculated using only $\mate_0$ and pre-computed values related to the $\beta_t$ sequence. This significantly speeds up the forward process by avoiding iterative calculations. Let $\alpha_t=1-\beta_t$ and $\bar{\alpha}t=\prod{t'=1}^T\alpha_{t'}$, the iterative formulation of $\mate_t$ can be simplified to an equation that depends only on $\mate_0$ and $\bar{\alpha}_t$:
\begin{align}
    \mate_t &= \sqrt{\alpha_t} \mate_{t-1} + \sqrt{\beta_t} \matn_1\nonumber\\
    &=\sqrt{\alpha_t}(\sqrt{\alpha_{t-1}} \mate_{t-2} + \sqrt{\beta_{t-1}} \matn_2 ) + \sqrt{\beta_t} \matn_1\nonumber\\
    &=\sqrt{\alpha_t\alpha_{t-1}}\mate_{t-2} + \sqrt{1-\alpha_t\alpha_{t-1}} \matn'_2\nonumber\\
    &=\sqrt{\bar{\alpha}_t} \mate_0 + \sqrt{1-\bar{\alpha}_t}{\matn}'_t
\end{align}
Here, $\matn$, $\matn'$ denote independent Gaussian distributions following $\mathcal{N}(0, \mati)$. Due to the addition rule of Gaussian distributions, the term $\sqrt{\alpha_t-\alpha_t\alpha_{t-1}}\matn_2+\sqrt{1-\alpha_t}\matn_1$ follows $\gauss(0,\sqrt{1-\alpha_t\alpha_{t-1}})$. By pre-calculating $\bar{\alpha}_t$ for $1\leq t\leq T$, $\mate_t$ can be efficiently obtained without recursion, facilitating random sampling for the diffusion step $t$.\\\vspace{-0.12in}

\noindent \textbf{Reverse Process}.
In the reverse process, our objective is to restore the social relation data in the hidden space from the noisy data $\mate_t$, where $t=1,\cdots,T$. Specifically, this process aims to estimate the following conditional probability using learnable neural networks:
\begin{align}
    \label{eq:gaus_output}
    p(\mate_{t-1}|\mate_t)=\gauss\left(\mate_{t-1}; \mathbf{\mu}_\theta(\mate_t, t),\mathbf{\Sigma}_\theta(\mate_t,t)\right)
\end{align}
Here, $\mathbf{\mu}_\theta(\cdot)$ and $\mathbf{\Sigma}_\theta(\cdot)$ represent neural networks with learnable parameters $\theta$ used to estimate the Gaussian distribution. We concatenate the $t$-step data vector with a time step-specific embedding as the input. The network consists of two fully-connected layers:
\begin{align}
    \mathbf{\mu}_\theta(\vece_t,t)=\textbf{FC}^2(\vece_t\| \textbf{h}_t),~~\textbf{FC}(\textbf{x})=\delta(\matw\vecx+\vecb)
\end{align}
In the above equations, $\textbf{FC}^2(\cdot)$ represents two consecutive fully-connected layers. $\vece_t\in\dmnr^d$ denotes a node embedding vector in the $t$-th diffusion step, and $\vech_t\in\dmnr^{d'}$ represents the embedding for the $t$-th time step. $\delta(\cdot)$, $\matw$, and $\vecb$ refer to the activation function, linear transformation, and bias vectors for the fully-connected layer.

\subsubsection{\bf Diffusion Loss Function}
The learnable denoising process of our hidden-space social diffusion is optimized by maximizing the evidence lower bound (ELBO) of the input social embeddings $\mate_0$. This ELBO term can be decomposed as follows:
\begin{align}
    &\log p(\vece_{0}) 
    =\log \int p\left(\vece_{0: T}\right) \mathrm{d} \vece_{1: T} %\nonumber\\
    =\log \mathbb{E}_{q(\vece_{1: T} | \vece_{0})}\frac{p(\vece_{0: T})}{q(\vece_{1: T} | \vece_{0})} \nonumber\\
    &\geq \underbrace{\mathbb{E}_{q\left(\vece_{1} | \vece_{0}\right)}\left[\log p_{\theta}\left(\vece_{0} | \vece_{1}\right)\right]}_{(\text {reconstruction term) }}-\underbrace{D_{\mathrm{KL}}\left(q\left(\vece_{T} | \vece_{0}\right) \| p\left(\vece_{T}\right)\right)}_{\text {(prior matching term) }} \nonumber\\
    &-\sum_{t=2}^{T} \underbrace{\mathbb{E}_{q\left(\vece_{t} | \vece_{0}\right)}\left[D_{\mathrm{KL}}\left(q\left(\vece_{t-1} | \vece_{t}, \vece_{0}\right) \| p_{\theta}\left(\vece_{t-1} | \vece_{t}\right)\right)\right]}_{\text {(denoising matching term) }}
\end{align}
Since the prior matching term is a constant, it can be omitted in the loss function for optimization. The denoising matching term aims to minimize the KL divergence between the true distribution $q(\vece_{t-1}|\vece_t,\vece_0)$ and our denoiser $p_\theta(\vece_{t-1}|\vece_t)$. Following previous works (Ho et al., 2020; Wang et al., 2023), we simplify the learning process by omitting the learning of the standard deviation network and assume that $\mathbf{\Sigma}_\theta(\vece_t,t)=\sigma^2(t)\mati$. The denoising matching term for the $t$-th time step can be defined as follows:
\begin{align}
    \mathcal{L}_t&=\mathbb{E}_{q\left(\vece_{t} | \vece_{0}\right)}\left[D_{\mathrm{KL}}\left(q\left(\vece_{t-1} | \vece_{t}, \vece_{0}\right) \| p_{\theta}\left(\vece_{t-1} | \vece_{t}\right)\right)\right]\nonumber\\
    &=\mathbb{E}_{q(\vece_t|\vece_0)}\left[ \frac{1}{2\sigma^2(t)} \left[\|\mathbf{\mu}_\theta(\vece_t,t)-\mathbf{\mu}(\vece_t,\vece_0,t)\|_2^2\right] \right]
\end{align}
$\mathbf{\mu}\theta(\vece_t,t)$ represents the output of our mean value predictor for an embedding vector $\vece_t$, and $\mathbf{\mu}(\vece_t,\vece_0,t)$ denotes the mean value for the true probability. These mean value terms can be decomposed into components related to $\vece_t$, $\vece_0$, and the output of a prediction network $\hat{\vece}\theta(\vece_t,t)$ for the real data $\vece_0$. As a result, $\loss_t$ can be expressed as:
\begin{align}
    \loss_t=\mathbb{E}_{q(\vece_t|\vece_0)}\left[\frac{1}{2}\left(\frac{\bar{\alpha}_{t-1}}{1-\bar{\alpha}_{t-1}} - \frac{\bar{\alpha}_t}{1-\bar{\alpha}_t} \right) \|\hat{\vece}_\theta(\vece_t,t)-\vece_0\|_2^2 \right]
\end{align}
The reconstruction term can be represented as the squared loss between the predicted value $\hat{\vece}\theta(\vece_1,1)$ and the real embedding vector $\vece_0$. This term is denoted as $\mathcal{L}^{'}_{t}$ and defined as follows:
\begin{align}
    \label{eq:squared_loss}
    \mathcal{L}^{'}_{t}=\mathbb{E}_{q(\vece_t|\vece_0)}\left[ \|\hat{\vece}_\theta(\vece_1,1)-\vece_0\|_2^2 \right]
\end{align}

\subsubsection{\bf Inference Process}
During the inference process, we focus on removing noise from the observed social data and utilize the denoised embeddings for making predictions. The social denoiser takes the encoded social embeddings $\mate^s$ as input, skipping the forward noise diffusion process. In the denoising step, we iteratively remove noise by updating $\hat{\vece}_{t-1} = \hat{{\mu}}_{\theta}(\hat{\vece}_{t}, t)$, where $\hat{{\mu}}_{\theta}(\hat{\vece}_t,t)$ is defined as follows:
\begin{align}
    \label{eq:infer_process}
    \hat{{\mu}}_{\theta}\left(\hat{\vece}_{t}, t\right)=\frac{\sqrt{\alpha_{t}\left(1-\bar{\alpha}_{t-1}\right)}}{1-\bar{\alpha}_{t}} \hat{\vece}_{t}+\frac{\sqrt{\bar{\alpha}_{t-1}\left(1-\alpha_{t}\right)}}{1-\bar{\alpha}_{t}} \hat{\vece}_{\theta}\left(\hat{\vece}_{t}, t\right)
\end{align}
We predict $\hat{\vece}_{0}$ based on $\hat{\vece}{t}$ and $t$, denoted as $\hat{\vece}_{\theta}$. We then use the derived $\hat{\vece}{0}$ for consecutive predictions. The inference procedure is outlined in Algorithm~\ref{alg:inference}.
\begin{algorithm}[t]
    \small
    \KwIn{The social tie embedding $\mate^{S}$. }
    \KwOut{the denoising embedding $\hat{\vece}_0$. }
    \ Set $\hat{\mate_t}$ = $\hat{\mate_0}$;\\
    \For{\textit{t} = $T$,...,1 }{
    $\hat{\vece}_{t-1} = \hat{\boldsymbol{\mu}}_{\theta}\left(\hat{\vece}_{t}, t\right) $ obtained from $\hat{\vece}_{t}$ and $\hat{\vece}_{\theta}(.)$ via  Eq~\ref{eq:infer_process};}
    \ return the denoising $\hat{\vece}_{0}$;\\
    \caption{{\bf Inference of our \model\ framework.} 
    \label{alg:inference}}
\end{algorithm}

\subsection{Prediction and Optimization}
Using the hidden-space social diffusion module, we combine the denoised social relation with the encoded interaction patterns to obtain final embeddings for predictions. This is done as follows:
\begin{align}
    \label{eq:fused_emb}
    \hat{r}_{u,v} = \tilde{\vece}_u^\top \vece_v^r,~~~~~\tilde{\vece}_u = \vece_u^r + \hat{\vece}_\theta(\vece_{u,t},t)
\end{align}
where $t$ represents a sampled diffusion step for user $u$. We optimize our \model\ using the predictions $\hat{r}_{u,v}$ and a combination of recommendation loss and diffusion loss functions:
\begin{align}
     \label{eq:total_loss}
    \loss=\sum_{u,v^+,v^-} -\log \sigma(\hat{r}_{u,v^+} - \hat{r}_{u,v^-}) + \lambda_1 \sum_{t} \loss'_t% + \lambda_2\|\mathbf{\Theta}\|_\text{F}^2
\end{align}
$(u,v^+,v^-)$ is a triplet sample for pairwise recommendation training \cite{rendle2009bpr}. The diffusion loss is calculated on sampled diffusion steps $t$ for embeddings. We also apply weight-decay regularization, with weight $\lambda_2$, to all trainable parameters $\mathbf{\param}$. The learning process of \model, including graph encoding, multi-step forward and back diffusion, and loss calculations.

\subsection{Model Complexity Analysis}
This section provides a comprehensive analysis of the time and space complexity of our \model\ with the social diffusion module. \\\vspace{-0.12in}

\noindent \textbf{Time Complexity}: Initially, \model\ performs graph-level information propagation on both the holistic collaborative graph $\graph_r$ and the social graph $\graph_s$. This process requires $\calo((|\sete_r|+|\sete_s|)\times d)$ calculations for message passing and $\calo((|\setu|+|\setv|)\times d^2)$ for embedding transformation. However, our social diffusion model operates exclusively on the current batch during each training step. Let $B$ be the number of user-item interaction pairs in each batch. The forward diffusion process costs $\calo(B\times d)$ computations, while the reverse process costs $\calo(B\times (d^2 +dd'))$. Theoretical analysis suggests that our \model\ achieves comparable time costs to common social recommendation methods based on GNNs.\\\vspace{-0.12in}

\noindent \textbf{Memory Complexity}: The graph encoding process of our \model\ model requires a similar number of parameters as conventional graph-based social recommenders. The hidden-space diffusion network employs $\calo(d^2+dd')$ parameters for the denoiser. In comparison, diffusion models operating on the original graph data typically require $\calo(|\setu|\times d)$ parameters for the trained denoising process. This distinction results in the denoiser of our \model\ having a smaller solution space, thereby alleviating optimization challenges.

\section{Evaluation}
\label{sec:eval}
In this section, we analyze the performance of our \model\ framework by exploring the following research questions (RQs):
\begin{itemize}[leftmargin=*]
\item \textbf{RQ1}: How does the performance of our \model\ model compare to various recommendation baseline methods?

\item \textbf{RQ2}: What are the effects of different designed modules in our \model\ framework on recommendation performance?

\item \textbf{RQ3}: How do different settings of key hyperparameters impact the recommendation accuracy of our \model\ method?

\item \textbf{RQ4}: What impact does the noise scale in the noise diffusion process of our \model\ have on the model's performance?

\item \textbf{RQ5}: How does the efficiency of \model\ compare to baselines?

\item \textbf{RQ6}: How effectively can our \model\ with the social diffusion model handle noisy user connections?

\item \textbf{RQ7}: Can our social relation denoiser provide explainability?

\end{itemize}

\begin{table}[t]
    \centering
    \caption{Statistics of experimental datasets.}
    \vspace{-0.15in}
    \label{tab:datasets}
    % \small
    % \resizebox{0.5\textwidth}{!}{%
        \begin{tabular}{l|c|c|c}
            \hline
            Dataset & Ciao & Yelp & Epinions \\ \hline
            \# of Users & 1925 & 99262 & 14680 \\
            \# of Items & 15053 & 105142 & 233261 \\
            \# of User-Item Interactions & 23223 & 672513 & 447312 \\
            \# of Social Interactions & 65084 & 1298522 & 632144 \\ \hline
        \end{tabular}%
    % }
    \vspace{-0.1in}
\end{table}

\subsection{Experimental Settings}
\subsubsection{\textbf{Experimental Datasets}}
\begin{figure}[t]
    \begin{minipage}[t]{0.15\textwidth}
		\centering
		\includegraphics[width=3cm]{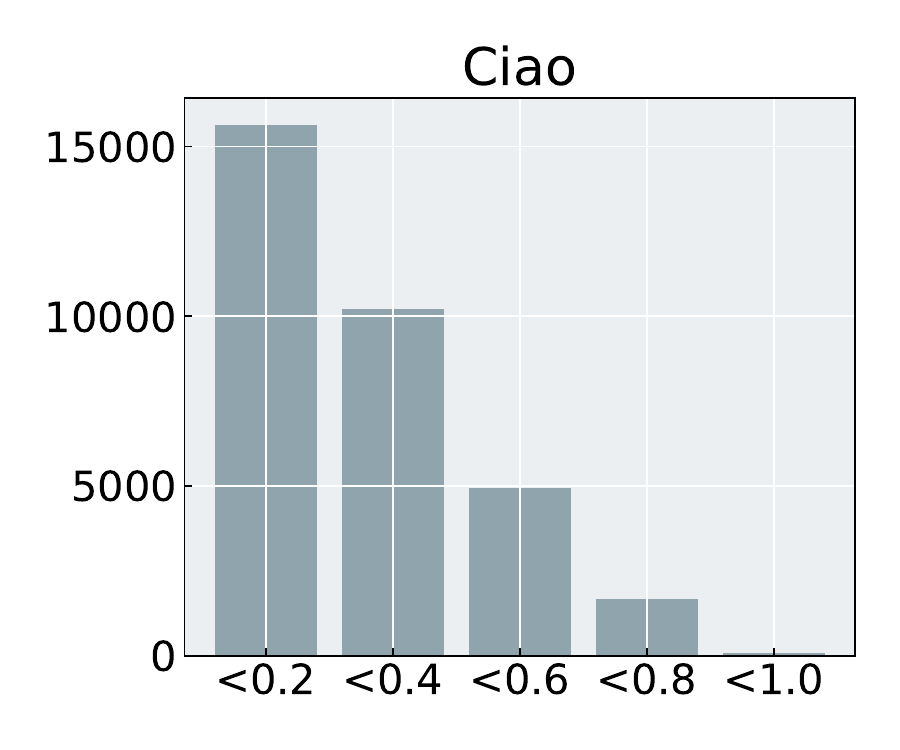}
	\end{minipage}
    \begin{minipage}[t]{0.15\textwidth}
		\centering
		\includegraphics[width=3cm]{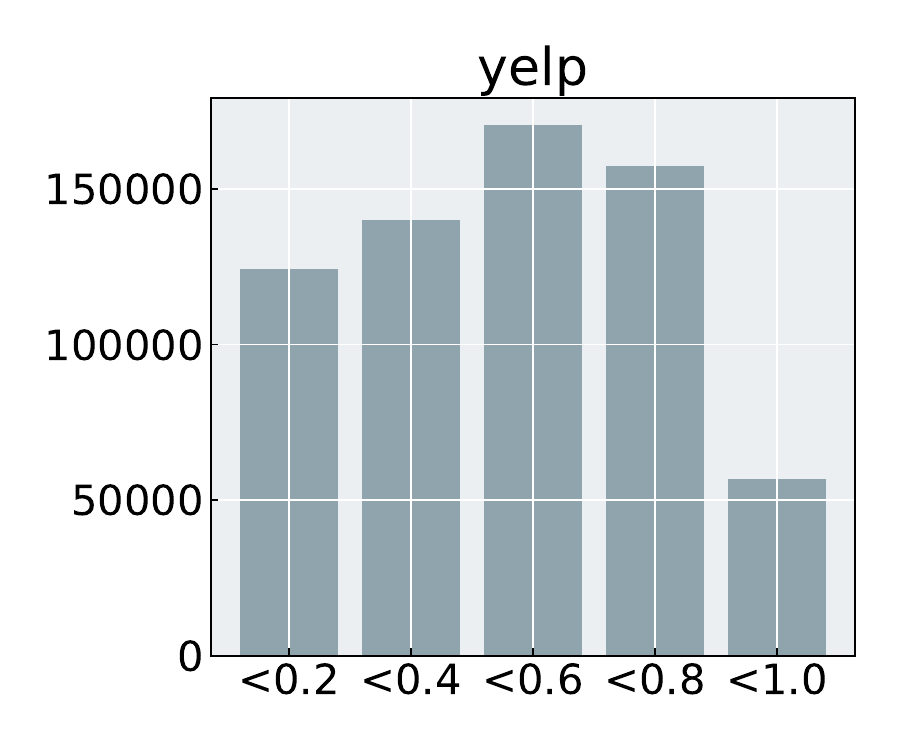}
	\end{minipage}
    \begin{minipage}[t]{0.15\textwidth}
		\centering
		\includegraphics[width=3cm]{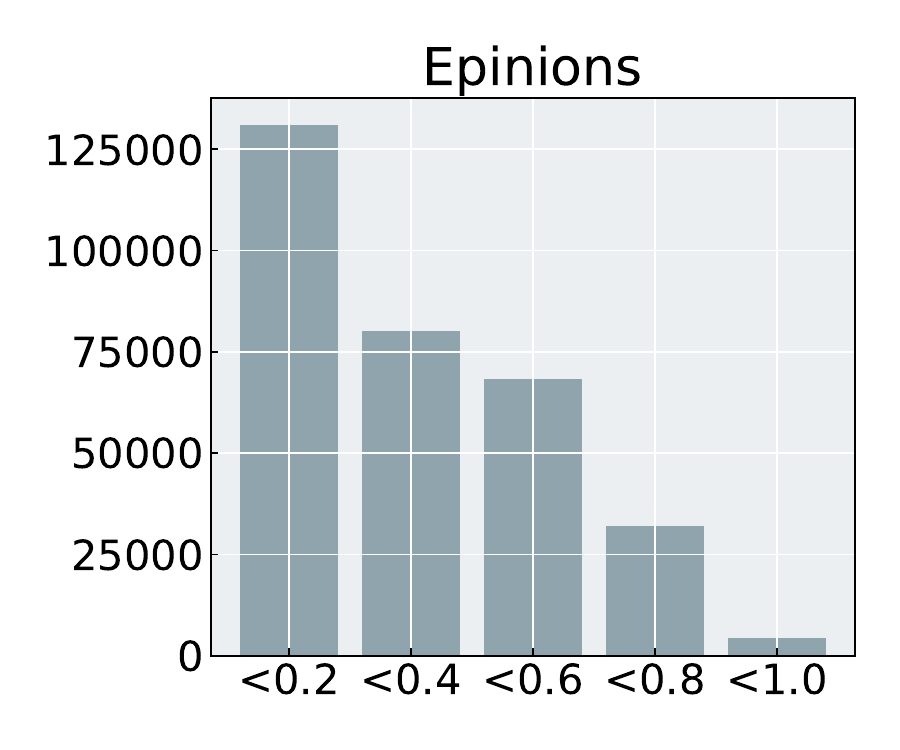}
	\end{minipage}
    \vspace{-0.15in}
    \caption{The distribution of social relation pairs across different datasets based on embedding similarity levels.}
    \vspace{-0.15in}
    \label{fig:noise_evidence}
    \vspace{-0.05in}
\end{figure}

We conducted experiments on three publicly available datasets from real-world commercial platforms: Yelp, Ciao, and Epinions. The user-item interaction data was based on users' review records, where an interaction $(u,v)$ exists if user $u$ reviewed item $v$. For user-wise social relationships, we established connections between users $(u, u')$ if user $u'$ is in the trust list of user $u$. Table~\ref{tab:datasets} presents the statistics of these datasets. Here is a summary of the dataset information:
\begin{itemize}[leftmargin=*]
\item \textbf{Yelp}: This data originates from Yelp and records user feedback on venues. It includes social relationships between users, emphasizing networks formed by individuals with shared interests.

\item \textbf{Ciao}: The Ciao dataset is sourced from the Ciao platform and captures user reviews and ratings across a wide range of products and services. It provides detailed information on social interactions among users, highlighting networks formed through shared preferences and engagements.

\item \textbf{Epinions}: This data collects user feedback on various items from the social network-based review platform Epinions \cite{fan2019graph}. It categorizes ratings from 1 to 5 into distinct interaction categories: negative, below average, neutral, above average, and positive.
\end{itemize}
We conducted an analysis to reveal the inconsistency between users' social relationships and their interaction patterns, indicating the presence of noise. By computing cosine similarity of social relation pairs' embeddings in the three datasets, we observed that a certain percentage of user pairs exhibit low-level similarity (cosine similarity < 0.2), suggesting the existence of noise in social relationships. Detailed results can be found in Table~\ref{fig:noise_evidence}.

\vspace{-0.05in}
\subsubsection{\textbf{Evaluation Protocols}}
In our experiments, we used a 7:1:2 ratio to create training, validation, and test sets for each dataset, following standard data partitioning criteria in graph-based recommender systems \cite{wang2019neural,he2020lightgcn}. To mitigate sampling bias, we employed an all-ranking protocol \cite{lin2022improving} to evaluate prediction accuracy for all items. The evaluation metrics used were \textit{Recall@N} and \textit{NDCG@N}, widely adopted in Top-N recommendations.

\begin{table*}[t]
    \centering
    \small
    \caption{Recommendation performance of all methods in terms of Recall@20 and NDCG@20.}
    \vspace{-0.15in}
    \label{tab:performance}
    \setlength{\tabcolsep}{1.1mm}
    % \resizebox{\textwidth}{!}{%
   \begin{tabular}{c|c|c|c|c|c|c|c|c|c|c|c|c||c|c}
        \hline
        Dataset & Metrics & TrustMF & SAMN & DiffNet & {GraphRec} & DGRec & NGCF & MHCN & KCGN & SMIN & {GDMSR} &DSL & \textbf{\model} & p-val \\ \hline\hline
        \multirow{4}{*}{Ciao} & Recall & 0.0539 & 0.0604 & 0.0528 & 0.0540& 0.0517 & 0.0559 & 0.0621 & 0.0602 & 0.0588 & 0.0560 & 0.0606 & \textbf{0.0712} & 4.047-12 \\ \cline{2-15} 
         & Imprv& 32.10\% & 17.88\% & 34.85\%& 31.85\% & 37.72\% & 27.37\% & 14.65\% & 18.27\% & 21.09\% & 27.14\%& 17.49\% & - & - \\ \cline{2-15} 
         & NDCG & 0.0343 & 0.0384 & 0.0328 & 0.0335 & 0.0319 & 0.0363 & 0.0378 & 0.0350 & 0.0354 & 0.0355 & 0.0389 & \textbf{0.0419} & 4.675-4 \\ \cline{2-15} 
         & Imprv& 22.16\% & 9.11\% & 27.74\% & 25.07\% & 31.35\% & 15.43\% & 10.85\% & 19.71\% & 18.36\% &18.03\% & 7.71\% & - & - \\ \hline
        \multirow{4}{*}{Yelp} & Recall & 0.0371 & 0.0403 & 0.0557 & 0.0419 & 0.0410 & 0.0450 & 0.0567 & 0.0460 & 0.0485 & 0.0513 & 0.0504 & \textbf{0.0597} & 4.288-10 \\ \cline{2-15} 
         & Imprv& 60.92\% & 48.14\% & 7.18\% &42.48\% & 45.61\% & 32.67\% & 5.29\% & 29.78\% & 23.09\% &16.37\% & 18.45\% & - & - \\ \cline{2-15} 
         & NDCG & 0.0193 & 0.0208 & 0.0292 & 0.0201 & 0.0209 & 0.0230 & 0.0292 & 0.0234 & 0.0251 & 0.0246 & 0.0259 & \textbf{0.0308} & 5.064-08 \\ \cline{2-15} 
         & Imprv& 59.59\% & 48.08\% & 5.48\% & 53.23\%& 47.37\% & 33.91\% & 5.48\% & 31.62\% & 22.71\% & 25.20\%& 18.92\% & - & - \\ \hline
        \multirow{4}{*}{Epinions} & Recall & 0.0265 & 0.0329 & 0.0384 & 0.0334& 0.0326 & 0.0353 & 0.0438 & 0.0370 & 0.0333 & 0.0368 & 0.0365 & \textbf{0.0460} & 1.117-08 \\ \cline{2-15} 
         & Imprv& 73.58\% & 39.82\% & 19.79\% & 37.72\% & 41.10\% & 30.31\% & 5.02\% & 24.32\% & 38.14\% & 25.00\%& 26.03\% & - & - \\ \cline{2-15} 
         & NDCG & 0.0195 & 0.0226 & 0.0273 & 0.0246 & 0.0236 & 0.0243 & 0.0321 & 0.0264 & 0.0228 &0.0241 & 0.0267 & \textbf{0.0336} & 4.200-08 \\ \cline{2-15} 
         & Imprv& 72.31\% & 48.67\% & 23.08\% & 36.59\%& 42.37\% & 38.27\% & 4.67\% & 27.27\% & 47.37\% & 39.42\%& 25.84\% & - & - \\ \hline
        \end{tabular}%
    % }
    \vspace{-0.05in}
\end{table*}

\subsubsection{\textbf{Compared Baselines}}
We compared our proposed model with 11 baseline methods representing diverse research approaches. The baseline methods used for comparison include:

\noindent\textbf{(i) \underline{Conventional and Attention-based Social Recommenders}}:

\noindent\
$\bullet$ \textbf{TrustMF} \cite{yang2016social}: Joint matrix factorization for user-item interaction and user-user trust matrices to enhance collaborative recommendations.\
$\bullet$ \textbf{SAMN} \cite{chen2019social}: Two-stage attention mechanism modeling social-aware relations between users and their social neighbors.\\\vspace{-0.12in}

\noindent{\textbf{(ii) \underline{Graph Collaborative Filtering Models}}}:

\noindent\
$\bullet$ \textbf{NGCF} \cite{wang2019neural}: A representative graph-enhanced collaborative filtering model that captures collaborative signals by propagating embeddings on GCN-layers using a user-item bipartite graph.\\\vspace{-0.12in}

\noindent{\textbf{(iii) \underline{GNN-based Social Recommender Systems}}}:

\noindent\
$\bullet$ \textbf{GraphRec} \cite{fan2019graph}: Introduces a graph attention network for attentive information propagation on the social network, merging social connection and user-item interaction information to enhance user representations.
$\bullet$ \textbf{DiffNet} \cite{wu2019neural}: Utilizes a layer-wise diffusion architecture to represent social relations through graph information propagation, capturing recursive social influences. $\bullet$ \textbf{GDMSR}\cite{quan2023robust}: Presents a robust graph-based denoising framework that effectively filters out noise, improving recommendation quality through preference guidance and relational modeling.\\\vspace{-0.12in}

\noindent{\textbf{(iv) \underline{Temporal-aware Social Recommendation Frameworks}}}:

\noindent\
$\bullet$ \textbf{DGRec} \cite{song2019session}: Combines RNNs with graph attention layers to capture dynamic user interests and social connections.\\\vspace{-0.12in}

\noindent{\textbf{(v) \underline{Knowledge-enhanced Social Recommender Systems}}}:

\noindent\
$\bullet$ \textbf{KCGN} \cite{huang2021knowledge}: Integrates interdependent knowledge among items and social influences among users within a multi-task learning framework for social recommendation.\\\vspace{-0.12in}

\noindent{\textbf{(vi) \underline{Self-Supervised Learning Social Recommenders}}}:

\noindent $\bullet$ \textbf{MHCN} \cite{yu2021self}: Self-supervised learning with multi-channel hypergraph neural networks to enhance model performance.\
$\bullet$ \textbf{SMIN} \cite{long2021social}: Proposes a meta-path-guided heterogeneous graph learning approach with self-supervised signals based on mutual information maximization to enhance training in social recommendation. $\bullet$ \textbf{DSL}\cite{wang2023denoised}: Introduces an adaptive self-supervision task for personalized social information denoising, preserving valuable social relationships for user preference modeling.

\vspace{-0.1in}
\subsubsection{\bf \emph{Hyperparameter Settings}}
We implement \model\ using PyTorch and optimize it with the Adam algorithm. The embeddings' dimensionality is tuned from ${8, 16, 32, 64}$. The learning rate ranges from ${0.001, 0.005, 0.01}$, and the batch size varies between 512 and 4096. The number of diffusion steps ranges from 10 to 200. Timestep embedding size is selected from ${4, 8, 16, 32}$. Other hyperparameter details can be found in our release code. For baselines, we use released code or implement them based on the original paper. Hyperparameters are optimized through grid search, with a standardized embedding dimension of 64. The batch size for Ciao is 2048, while for Yelp and Epinions, it is 4096. GNN models use 1 to 3 propagation layers for optimal performance across baselines.

\vspace{-0.1in}
\subsection{Overall Performance Comparison (RQ1)}
\begin{table*}[t]
    \centering
    \small
    % \footnotesize
    \caption{Recommendation performance with varying top-N settings, in terms of Recall@N and NDCG@N.}
    \vspace{-0.15in}
    \label{tab:topn}
    % \caption{PERFORMANCE EVALUATION WITH VARYING TOP-N IN TERMS OF Recall@N AND NDCG@N.}
    % \resizebox{\textwidth}{!}{%
    \begin{tabular}{c|cc|cc|cc|cc|cc|cc}
    \hline
    \multirow{2}{*}{Model} & \multicolumn{2}{c|}{Ciao@ 10} & \multicolumn{2}{c|}{Ciao@ 40} & \multicolumn{2}{c|}{Yelp@ 10} & \multicolumn{2}{c|}{Yelp@ 40} & \multicolumn{2}{c|}{Epinions @ 10} & \multicolumn{2}{c}{Epinions @ 40} \\ \cline{2-13} 
     & \multicolumn{1}{c|}{Recall} & NDCG & \multicolumn{1}{c|}{Recall} & NDCG & \multicolumn{1}{c|}{Recall} & NDCG & \multicolumn{1}{c|}{Recall} & NDCG & \multicolumn{1}{c|}{Recall} & NDCG & \multicolumn{1}{c|}{Recall} & NDCG \\ \hline\hline
    TrustMF & \multicolumn{1}{c|}{0.0341} & 0.0289 & \multicolumn{1}{c|}{0.0796} & 0.0416 & \multicolumn{1}{c|}{0.0224} & 0.0149 & \multicolumn{1}{c|}{0.0606} & 0.0254 & \multicolumn{1}{c|}{0.0165} & 0.0163 & \multicolumn{1}{c|}{0.0394} & 0.0236 \\ \hline
    SAMN & \multicolumn{1}{c|}{0.0345} & 0.0289 & \multicolumn{1}{c|}{0.0801} & 0.0429 & \multicolumn{1}{c|}{0.0289} & 0.0195 & \multicolumn{1}{c|}{0.0700} & 0.0308 & \multicolumn{1}{c|}{0.0193} & 0.0181 & \multicolumn{1}{c|}{0.0496} & 0.0280 \\ \hline
    DiffNet & \multicolumn{1}{c|}{0.0328} & 0.0271 & \multicolumn{1}{c|}{0.0780} & 0.0397 & \multicolumn{1}{c|}{0.0381} & 0.0247 & \multicolumn{1}{c|}{0.0739} & 0.0312 & \multicolumn{1}{c|}{0.0238} & 0.0227 & \multicolumn{1}{c|}{0.0587} & 0.0335 \\ \hline
    {GraphRec} & \multicolumn{1}{c|}{0.0322} & 0.0266 & \multicolumn{1}{c|}{0.0838} & 0.0420 & \multicolumn{1}{c|}{0.0233} & 0.0144 & \multicolumn{1}{c|}{0.0711} & 0.0277 & \multicolumn{1}{c|}{0.0207} & 0.0206 & \multicolumn{1}{c|}{0.0521} & 0.0304 \\ \hline
    DGRec & \multicolumn{1}{c|}{0.0296} & 0.0254 & \multicolumn{1}{c|}{0.0733} & 0.0381 & \multicolumn{1}{c|}{0.0245} & 0.0158 & \multicolumn{1}{c|}{0.0656} & 0.0272 & \multicolumn{1}{c|}{0.0197} & 0.0194 & \multicolumn{1}{c|}{0.0517} & 0.0293 \\ \hline
    NGCF & \multicolumn{1}{c|}{0.0366} & 0.0301 & \multicolumn{1}{c|}{0.0804} & 0.0343 & \multicolumn{1}{c|}{0.0276} & 0.0177 & \multicolumn{1}{c|}{0.0711} & 0.0297 & \multicolumn{1}{c|}{0.0217} & 0.0206 & \multicolumn{1}{c|}{0.0550} & 0.0304 \\ \hline
    MHCN & \multicolumn{1}{c|}{0.0343} & 0.0286 & \multicolumn{1}{c|}{0.0929} & 0.0473 & \multicolumn{1}{c|}{0.0342} & 0.0225 & \multicolumn{1}{c|}{0.0890} & 0.0377 & \multicolumn{1}{c|}{0.0272} & 0.0272 & \multicolumn{1}{c|}{0.0674} & 0.0321 \\ \hline
    KCGN & \multicolumn{1}{c|}{0.0360} & 0.0263 & \multicolumn{1}{c|}{0.0926} & 0.0448 & \multicolumn{1}{c|}{0.0284} & 0.0182 & \multicolumn{1}{c|}{0.0702} & 0.0295 & \multicolumn{1}{c|}{0.0221} & 0.0219 & \multicolumn{1}{c|}{0.056} & 0.0325 \\ \hline
    SMIN & \multicolumn{1}{c|}{0.0326} & 0.0275 & \multicolumn{1}{c|}{0.0813} & 0.0453 & \multicolumn{1}{c|}{0.0316} & 0.0198 & \multicolumn{1}{c|}{0.0768} & 0.0312 & \multicolumn{1}{c|}{0.0203} & 0.0186 & \multicolumn{1}{c|}{0.0531} & 0.0289 \\ \hline
    {GDMSR} & \multicolumn{1}{c|}{0.0340} & 0.276 & \multicolumn{1}{c|}{0.0804} & 0.0409 & \multicolumn{1}{c|}{0.0369} & 0.0196 & \multicolumn{1}{c|}{0.0744} & 0.0293 & \multicolumn{1}{c|}{0.0226} & 0.0206 & \multicolumn{1}{c|}{0.0536} & 0.0291 \\ \hline \hline 
    
    DSL & \multicolumn{1}{c|}{0.0412} & 0.0329 & \multicolumn{1}{c|}{0.0873} & 0.0473 & \multicolumn{1}{c|}{0.0315} & 0.0203 & \multicolumn{1}{c|}{0.0786} & 0.0332 & \multicolumn{1}{c|}{0.0229} & 0.0226 & \multicolumn{1}{c|}{0.0594} & 0.0338 \\ \hline \hline 
    \textbf{\model} & \multicolumn{1}{c|}{\textbf{0.0457}} & \textbf{0.0361} & \multicolumn{1}{c|}{\textbf{0.1023}} & \textbf{0.0535} & \multicolumn{1}{c|}{\textbf{0.0391}} & \textbf{0.0249} & \multicolumn{1}{c|}{\textbf{0.0941}} & \textbf{0.0394} & \multicolumn{1}{c|}{\textbf{0.0282}} & \textbf{0.0275} & \multicolumn{1}{c|}{\textbf{0.0696}} & \textbf{0.0343} \\ \hline 
    \end{tabular}%
    % }
    \vspace{-0.1in}
\end{table*}

We compare the overall recommendation performance of our \model\ with baselines. The comparison results are presented in Table~\ref{tab:performance} for top-20 evaluation and Table~\ref{tab:topn} for varying top-N evaluation. Based on these results, we draw the following conclusions.\\\vspace{-0.12in}

\noindent $\bullet$ \textbf{Superiority of \model}. Our \model\ consistently outperforms state-of-the-art baselines, demonstrating superior recommendation accuracy. T-tests confirm the statistical significance of our results across all datasets and evaluation metrics. The performance advantage of \model\ remains consistent across different top-N settings (Table~\ref{tab:topn}). Our diffusion-based social relation denoising module removes irrelevant and false information, allowing \model\ to effectively mine valuable social ties for enhanced recommendations. \\\vspace{-0.12in}

\noindent $\bullet$ \textbf{Negative Impact of Noisy Social Information}. Some social recommendation methods, such as DGRec, DiffNet, and GraphRec, perform worse than the social information-agnostic method NGCF. This suggests that social connections can have a negative influence on user-item relation modeling due to false or irrelevant components. Our \model\ framework addresses this issue by denoising social information and consistently outperforms the baseline model GDMSR. It effectively filters out noise from social connections and identifies meaningful and influential social ties, accurately encoding user preferences for precise recommendations. \\\vspace{-0.12in}

\noindent $\bullet$ \textbf{Advantages of diffusion-based supervision augmentation}. Baseline methods incorporating self-supervised learning (SSL) consistently outperform other approaches in recommendation performance. Methods like MHCN, KCGN, and SMIN utilize variations of the local-global infomax technique, while DSL employs a prediction alignment SSL task. This highlights the positive impact of auxiliary supervision signals in addressing data deficiency challenges in social recommendation, such as noise and sparsity. In contrast, our \model\ introduces a multi-step denoising method based on the diffusion model, generating a larger number of supervision signals at different noise levels. This robust denoising capability leads to superior recommendation performance, surpassing the baselines.

\vspace{-0.05in}
\subsection{Model Ablation Study (RQ2)}
In this section, we investigate the influence of different sub-modules in our \model\ framework through an ablation study. We evaluate the performance of several variants obtained by removing or replacing essential modules. The following ablated models are compared:\\\vspace{-0.12in}

\noindent $\bullet$ \textbf{-D}: Removes the holistic diffusion module, retaining only the social and user-item relation learning GNN.

\noindent $\bullet$ \textbf{-S}: Does not utilize social information. Instead, it solely relies on the user-item interaction graph to make recommendations.

\noindent $\bullet$ \textbf{DAE}: Replaces the diffusion-based denoising module of \model\ with a denoising autoencoder. This DAE-based denoising module is trained to reconstruct randomly masked user representations.\vspace{-0.05in}

\begin{figure}[h]
    \vspace{-0.05in}
    \begin{minipage}[t]{0.15\textwidth}
		\centering
		\includegraphics[width=2.8cm]{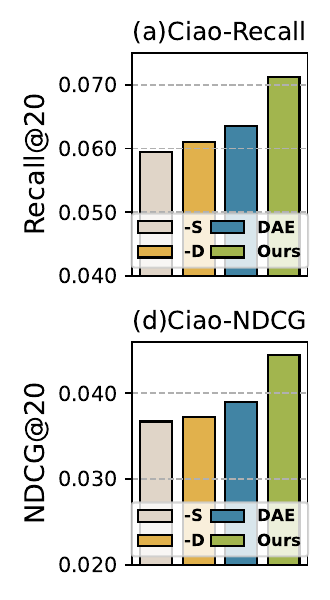}
	\end{minipage}
    \begin{minipage}[t]{0.15\textwidth}
		\centering
		\includegraphics[width=2.8cm]{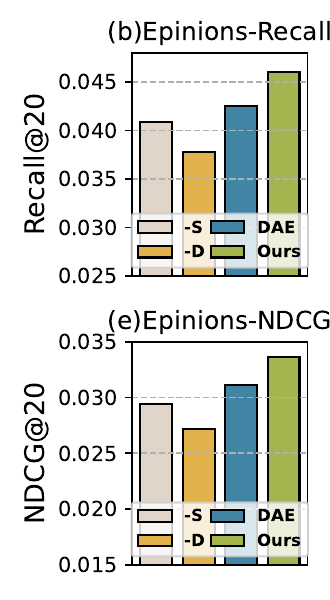}
	\end{minipage}
    \begin{minipage}[t]{0.15\textwidth}
		\centering
		\includegraphics[width=2.8cm]{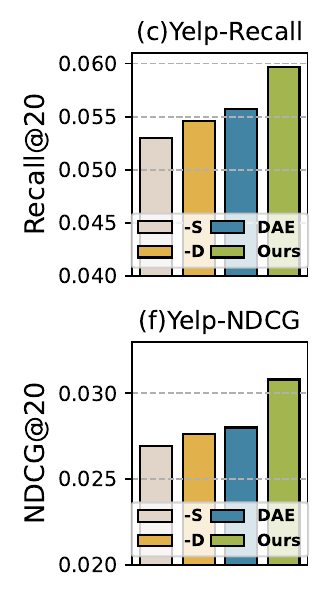}
	\end{minipage}
    \vspace{-0.2in}
    \caption{Ablation studies on Yelp, Ciao, and Epinions datasets for different sub-modules in our proposed \model\ framework, measuring Recall@20 and NDCG@20.}
    \label{fig:ablation}
    \vspace{-0.15in}
\end{figure}

We evaluate these variants on the Ciao and Yelp datasets using the top-20 recommendation setting. The results, depicted in Figure~\ref{fig:ablation}, consistently demonstrate that \model\ outperforms all four variants. These findings strongly support the effectiveness of the social relation learning and diffusion-based denoising components in our model. Notable observations from our analysis include:\\\vspace{-0.12in}

\noindent $\bullet$ (1) Removing the diffusion module (\textit{-D}) leads to significant performance degradation, highlighting the denoising function's effectiveness provided by our latent feature-level diffusion model.\\\vspace{-0.12in}

\noindent $\bullet$ (2) Comparing the \textit{-S} variant to \model\ highlights the significant improvement obtained by incorporating users' social context information in user preference learning. However, subgraphs (b) and (e) suggest that in the presence of noisy social information in the epinions dataset, the \textit{-S} variant may outperform the \textit{-D} variant.\\\vspace{-0.12in}

\noindent $\bullet$ (3) The suboptimal performance of the \textit{DAE} variant showcases the superior denoising ability of our designed diffusion module compared to vanilla denoising techniques. \model\ effectively models complex representation distributions by gradually learning each denoising transition step from $t$ to $t-1$ through shared neural networks, enhancing noise reduction in latent features.

\vspace{-0.1in}
\subsection{Impact of Hyperparameters (RQ3)}
\begin{figure*}[t]
    \centering
    \includegraphics[width=\textwidth]{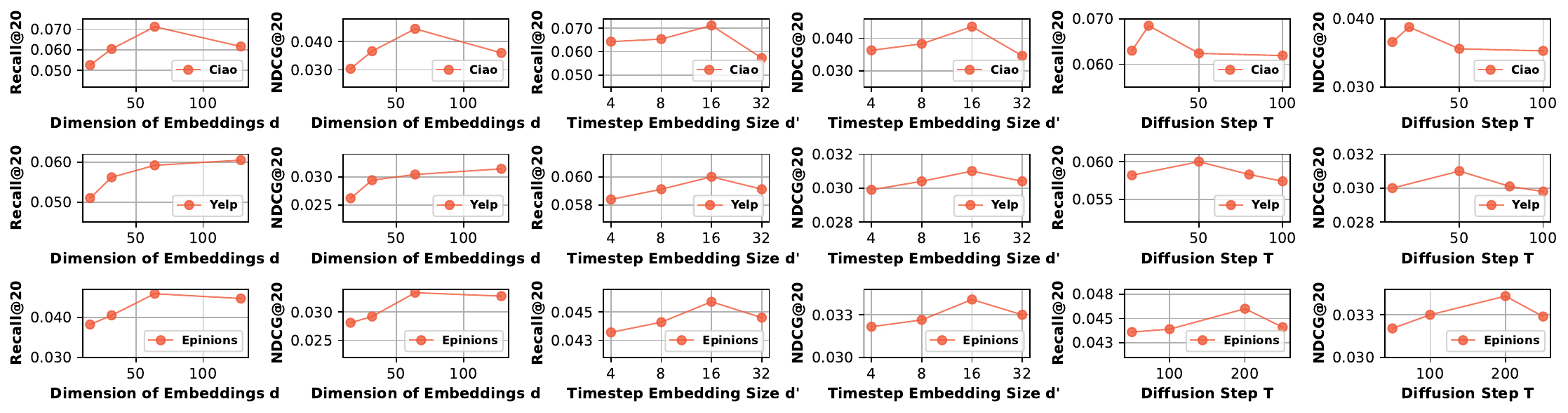}
    \vspace{-0.25in}
    \caption{Hyperparameter study on important parametric configurations of \model, in terms of Recall@20 and NDCG@20.}
    \vspace{-0.15in}
    \label{fig:hyper}
\end{figure*}

This section investigates the impact of crucial hyperparameters on model performance: the dimensionality of hidden representations ($d$), the dimensionality of time step embeddings ($d'$), and the maximum number of diffusion steps ($T$). Evaluation is conducted on all three experimental datasets, measuring Recall@20 and NDCG@20 metrics. The results, shown in Figure~\ref{fig:hyper}, are analyzed as follows:\\\vspace{-0.12in}

\noindent $\bullet$ \textbf{Embedding dimensionality} $d$: Increasing $d$ improves performance, except for the Ciao and Epinions datasets where larger values lead to slight degradation due to overfitting. \\\vspace{-0.12in}

\noindent $\bullet$ \textbf{Time step embedding size} $d'$: Larger dimensions enhance the positive impact of diffusion steps on denoising, improving performance. However, excessively high dimensions hinder the model's diffusion ability, resulting in decreased performance.\\\vspace{-0.12in}

\noindent $\bullet$ \textbf{Maximum diffusion steps} $T$: Increasing $T$ enhances performance with more diffusion steps. However, extremely large values damage social information and reduce denoising effectiveness.

\vspace{-0.1in}
\subsection{Impact of Noise Scale (RQ4)}
\begin{figure}[h!]
    \centering
    \includegraphics[width=\columnwidth]{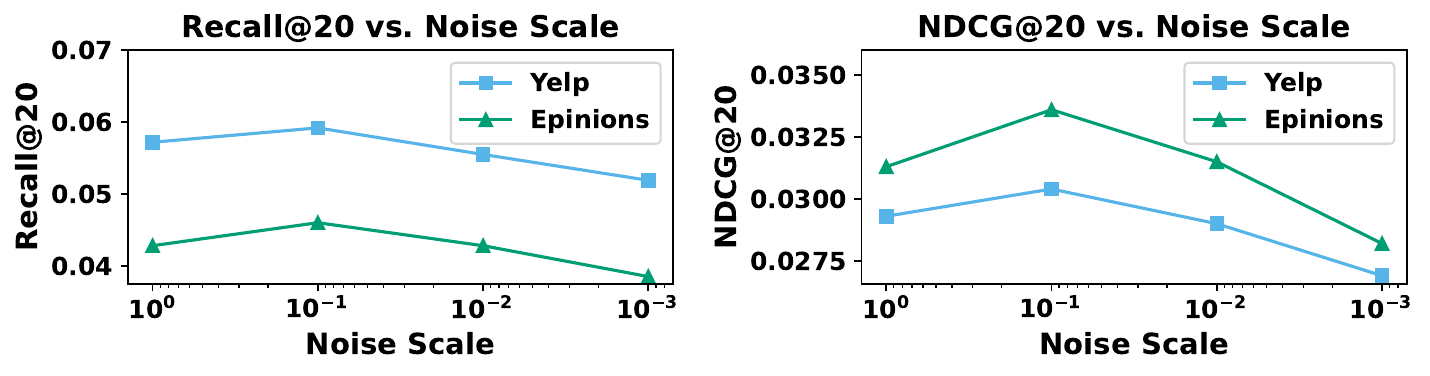}
    \vspace{-0.2in}
    \caption{Impact of noise scale over model performance.}
    \label{fig:noise_scale}
    \vspace{-0.15in}
\end{figure}

This section investigates the impact of the noise scale factor ($\tau$) on our noising process. By adjusting the minimum noise ($\bar{s}{min}$) and maximum noise ($\bar{s}{max}$) in the noise scheduler to $\tau\cdot\bar{s}{min}$ and $\tau\cdot\bar{s}{max}$, respectively, we examine the model's performance with different noise scale values of ${1, 0.1, 0.01, 0.001}$. The results, depicted in Figure~\ref{fig:noise_scale}, reveal the following insights:

\noindent $\bullet$ Increasing the noise scale effectively improves model performance, showcasing the effectiveness of our denoising mechanism within our proposed \model\ framework.\\\vspace{-0.12in}

\noindent $\bullet$ Further increasing the noise scale beyond a certain threshold leads to a decline in performance. This effect is particularly pronounced in sparsely populated datasets like Yelp and Epinion.

We attribute this decline to excessive noise obscuring the intrinsic personalized data, thereby hindering the model's ability to retain and process essential individualized information.

\vspace{-0.05in}
\subsection{Training Efficiency Study (RQ5)}
\begin{figure}[t]
    \centering
   \begin{minipage}[t]{0.48\columnwidth}
        \centering
        \includegraphics[width=\textwidth]{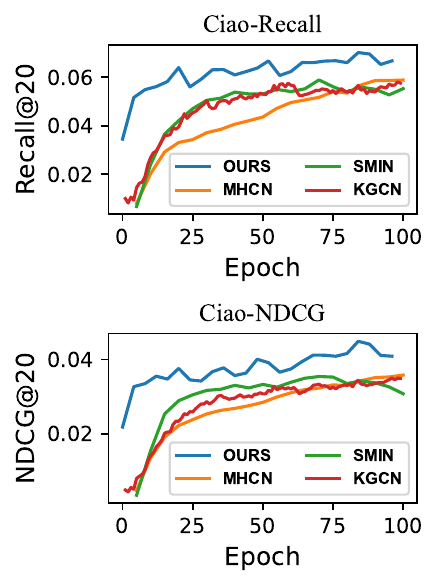}
    \end{minipage}
    \begin{minipage}[t]{0.48\columnwidth}
        \centering
        \includegraphics[width=\textwidth]{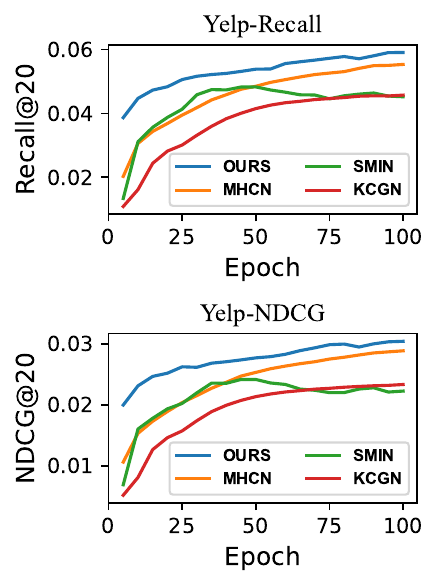}
    \end{minipage}
    \vspace{-0.2in}
    \caption{Model performance by training epochs on the Ciao and Yelp test sets, measured by Recall@20 and NDCG@20.}
    \label{fig:efficiency}
    \vspace{-0.25in}
\end{figure}

This section optimizes the efficiency of our \model\ compared to baseline models (MHCN, SMIN, and KCGN) on the Ciao and Yelp datasets. Using an A40 graphics card with 48GB of GPU memory, we compared the time costs of these baselines (Table~\ref{tab:time}). Our \model\ demonstrates significant efficiency advantages in both training and testing. For each training epoch, we evaluated and recorded the test set performance to analyze improvements (Figure~\ref{fig:efficiency}). \\\vspace{-0.12in}

\noindent $\bullet$ \textbf{Training efficiency of \model}: Our \model\ consistently outperforms the baselines in training efficiency, benefiting from effective denoising diffusion for accelerated optimization. \\\vspace{-0.12in}

\noindent $\bullet$ \textbf{Limitations of baselines}: SMIN shows overfitting effects, potentially due to reliance on metagraphs, limiting generalization. MHCN achieves high final performance but converges slower due to its complex hypergraph structure. In contrast, our \model\ benefits from a compact neural architecture without handcrafted priors, enabling faster optimization with auxiliary signals. \\\vspace{-0.12in}

\noindent $\bullet$ \textbf{Fluctuations on Ciao data}: In comparing convergence curves, significant recommendation performance fluctuations are observed on the smaller Ciao dataset, indicating training instability.

\subsection{Further Exploration of Anti-Noise Capacity with our \model\ Framework (RQ6)}
\begin{figure}[t]
    \centering
    \includegraphics[width=8cm]{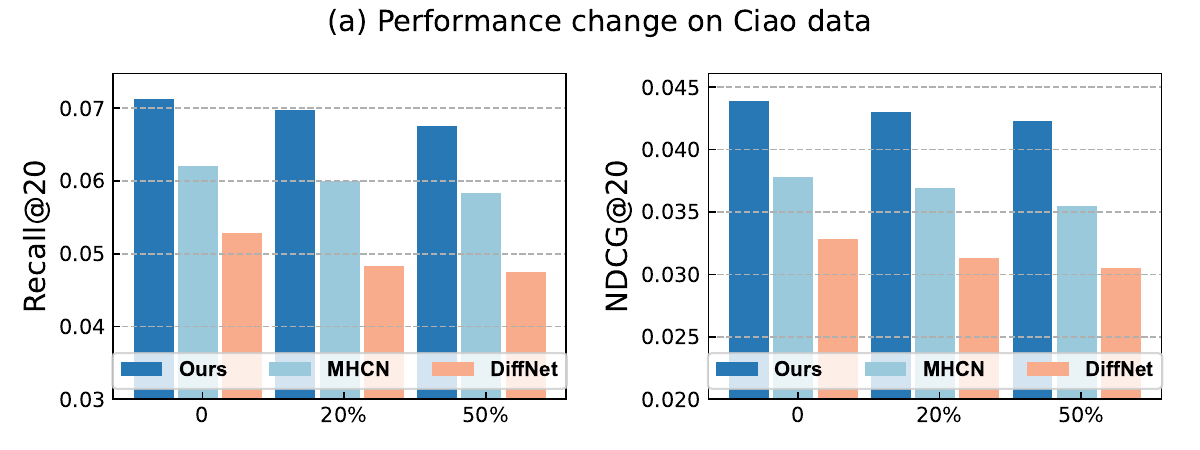}
    \includegraphics[width=8cm]{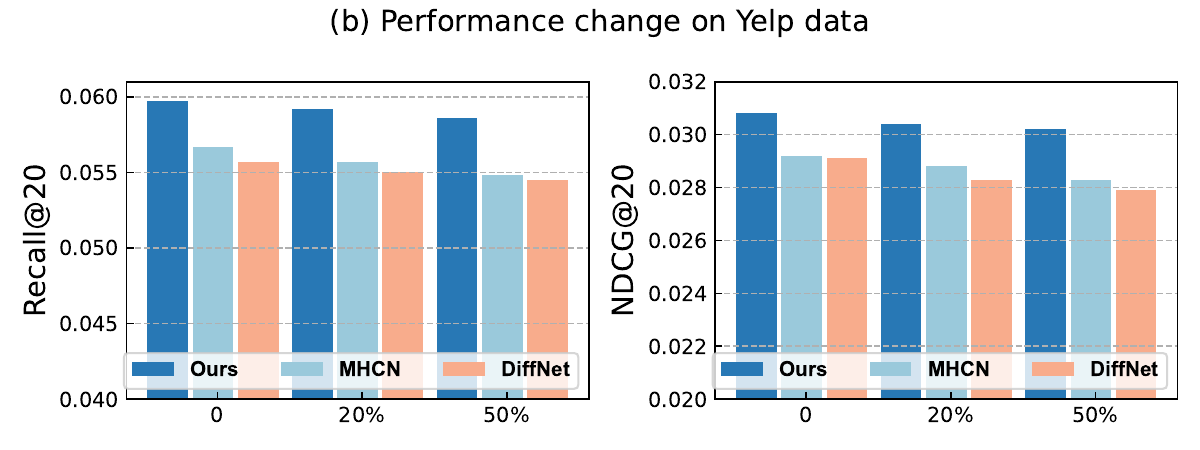}
    \includegraphics[width=8cm]{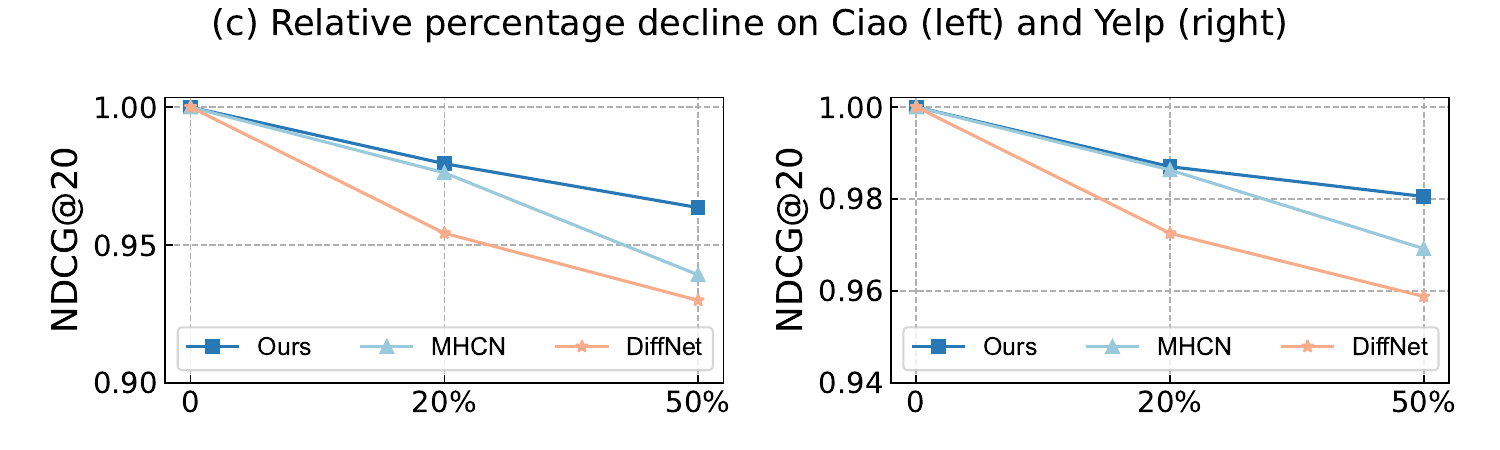}
    \vspace{-0.1in}
    \caption{Investigating the Influence of Different Noise Ratios on Performance Degradation.}
    \label{fig:noise}
    \vspace{-0.15in}
\end{figure}

We evaluate the robustness of \model\ in the presence of data noise by introducing random fake edges to replace varying percentages of genuine social connections in the user-user graph. The model is then retrained using the corrupted graph and evaluated on the true test set. Specifically, we analyze the effects of replacing 0\%, 20\%, and 50\% of the social relations with noise signals. Comparing the performance of \model\ with MHCN and DiffNet, the results in Figure~\ref{fig:noise} (a) and (b) show the original evaluation outcomes, while (c) illustrates the relative performance change in NDCG. Based on these results, the following observations are made:

\noindent $\bullet$ \textbf{Advantageous robustness of \model}: Our \model\ model outperforms the baselines with a smaller performance drop, showcasing its superior denoising capabilities in social recommendation.

\noindent $\bullet$ \textbf{Denoising effect of vanilla SSL}: The MHCN model shows promise in denoising, but it falls short compared to our \model\ model. This highlights that general-purpose self-supervised learning tasks may not effectively address the specific denoising requirements of social recommendation.

\noindent $\bullet$ \textbf{Higher noise ratio in Ciao dataset}: The Ciao dataset demonstrates a larger performance drop, suggesting a higher noise ratio in comparison to other datasets.

\begin{figure*}[t]
    \centering
    \includegraphics[width=0.9\textwidth]{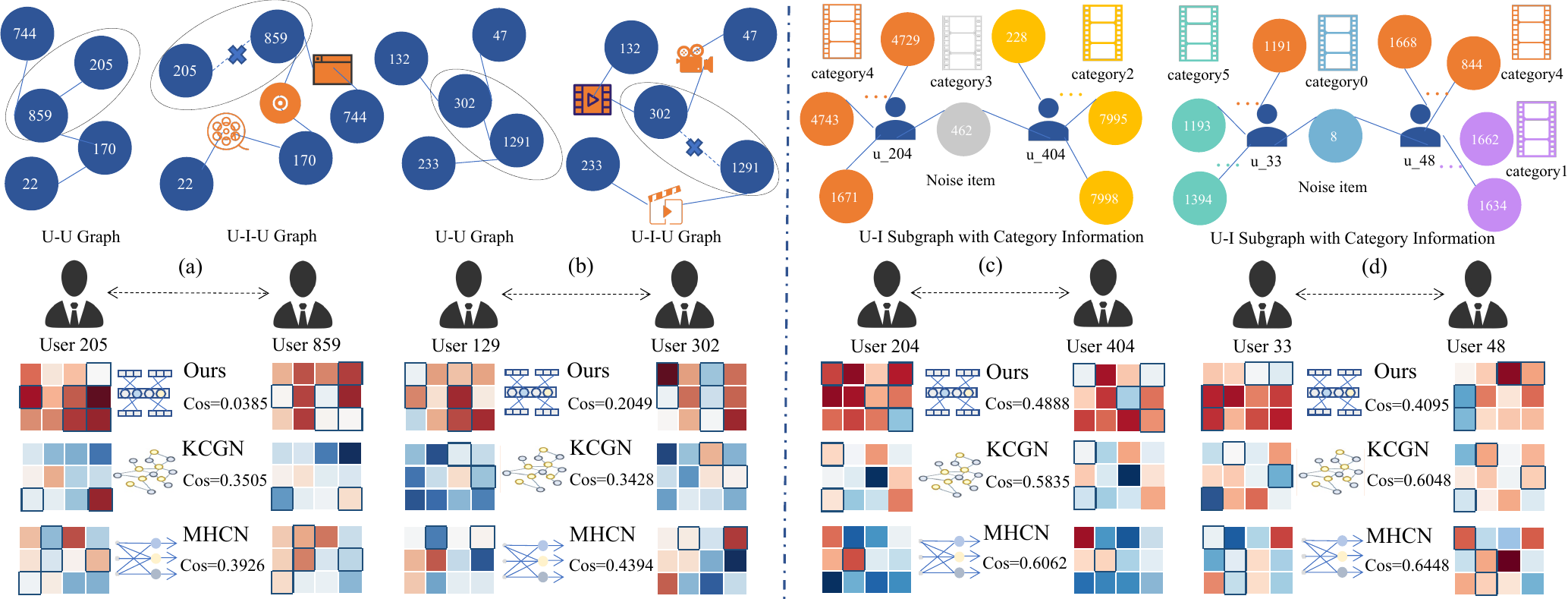}
    \vspace{-0.15in}
    \caption{Case study for the user relation recalibration effect of our social denoiser \model.}
    \label{fig:case_study}
    \vspace{-0.15in}
\end{figure*}

\begin{table}[]
    \centering
    \scriptsize
    \caption{{The running time (in seconds) per epoch for different models is compared on diverse evaluation datasets.}}
    \vspace{-0.1in}
    \label{tab:time}
    \resizebox{0.48\textwidth}{!}{%
    \begin{tabular}{|c|ccc|ccc|}
    \hline
    \multirow{2}{*}{Model} & \multicolumn{3}{c|}{Training} & \multicolumn{3}{c|}{Testing} \\ \cline{2-7} 
     & \multicolumn{1}{c|}{ciao} & \multicolumn{1}{c|}{yelp} & epinions & \multicolumn{1}{c|}{ciao} & \multicolumn{1}{c|}{yelp} & epinions \\ \hline
    MHCN & \multicolumn{1}{c|}{0.46} & \multicolumn{1}{c|}{28.51} & 9.98 & \multicolumn{1}{c|}{0.33} & \multicolumn{1}{c|}{44.39} & 22.74 \\ \hline
    KCGN & \multicolumn{1}{c|}{0.33} & \multicolumn{1}{c|}{75.84} & 63.11 & \multicolumn{1}{c|}{0.25} & \multicolumn{1}{c|}{31.23} & 13.61 \\ \hline
    SMIN & \multicolumn{1}{c|}{0.95} & \multicolumn{1}{c|}{60.86} & 66.15 & \multicolumn{1}{c|}{1.25} & \multicolumn{1}{c|}{35.54} & 27.42 \\ \hline
    Ours & \multicolumn{1}{c|}{0.19} & \multicolumn{1}{c|}{4.23} & 2.78 & \multicolumn{1}{c|}{0.31} & \multicolumn{1}{c|}{27.91} & 13.54 \\ \hline
    \end{tabular}%
    }
    \vspace{-0.2in}
\end{table}

\subsection{Case Study (RQ7)}
This section explores the denoising effect of \model\ on specific user/item cases. Four subgraph cases are illustrated in Figure~\ref{fig:case_study}, highlighting the need for denoising. The baseline methods, KCGN and MHCN, fail to identify false social connections, resulting in high cosine scores for the incorrect social neighbors. In contrast, our proposed \model\ effectively recognizes these noise instances, yielding significantly lower similarity scores and producing distinct embeddings for falsely-connected users. These findings demonstrate the superior noise elimination capability of \model\ across different noise situations.

Two additional cases are presented, involving user pairs sharing interactions with items that significantly differ in category from other items the users interact with. These isolated interactions are likely to be noisy items, rendering the associated social links noisy as well. Once again, \model\ successfully identifies and eliminates the noise, assigning lower similarity scores and generating more distinct embeddings for false social neighbors. These cases further exemplify the denoising effectiveness of our \model\ approach.

\vspace{-0.05in}
\section{Related Work}
\label{sec:relate}

\noindent $\bullet$ \textbf{Social-aware Recommender Systems}. Deep learning-based social recommender systems, such as DiffNet \cite{wu2019neural}, RecoGCN \cite{xu2019relation}, and KCGN \cite{huang2021knowledge}, leverage Graph Neural Networks to effectively model the connections between users and items. Approaches like SAMN \cite{chen2019social} and GraphRec \cite{fan2019graph} further enhance social recommendation by incorporating attention mechanisms to differentially differentiate the influence levels among users. More recent self-supervised learning (SSL) based methods, including MHCN \cite{yu2021self}, SMIN \cite{long2021social}, SDCRec \cite{du2022socially}, and DcRec \cite{wu2022disentangled}, have shown promising and encouraging results in bolstering social recommenders through innovative SSL-based data augmentation techniques. In contrast to these existing established approaches, our proposed method, \model, takes a unique and distinctive stance by concentrating on denoising relation learning in social recommender systems with the power of diffusion models. \\\vspace{-0.12in}

\noindent $\bullet$ \textbf{Recommendation with Graph Neural Networks}.
Graph neural networks have achieved state-of-the-art performance in recommendation scenarios \cite{yan20222, higpt2024, zhou2023layer}. Ealier works like NGCF \cite{wang2019neural} and Pinsage \cite{ying2018graph} introduced higher-order connectivity extraction using graph convolutional networks (GCNs). LightGCN \cite{he2020lightgcn} simplified training by removing non-linear activations, while DGCF \cite{wang2020disentangled} focused on intent-aware modeling through graph disentangling. Studies have also incorporated auxiliary information, such as multi-modal \cite{wei2019mmgcn} and knowledge \cite{wang2019kgat} data, to further enhance recommendation performance. Recent advancements in graph self-supervised learning include SGL \cite{wu2021self}, AutoCF \cite{autocfwww2023}, and NCL \cite{lin2022improving}, which have demonstrated promising results. In contrast, our work takes a unique approach by enhancing the denoising capabilities of recommenders through the development of an auxiliary task for social recommendation, leveraging a diffusion-based paradigm. \\\vspace{-0.12in}

\noindent $\bullet$ \noindent \textbf{Generative Models for Recommendation}.
Generative models, such as GANs~\cite{wang2017irgan} and VAEs~\cite{Yu_Zhang_Cao_Xia}, have gained prominence in recommender systems for data generation to enhance preference modeling~\cite{liu2021interest,Wei_Wang_Nie_Li_Wang_Chua_2022}. Some approaches focus on generating synthetic user-item interaction data to address data sparsity and cold start issues. GAN-based methods~\cite{Chen_Wang_Huang_Huang_Xu_Lin_Inc_Li, jin2020sampling, Wang_Ye_Chen_Zhang_Wang_Zou_Liu_Wang_Wang_Generative} use adversarial training to mimic real user behaviors. VAE-based generative recommendation models~\cite{liang2018variational, Ma_Zhou_Cui_Yang_Zhu_2019} employ variational autoencoders to make accurate predictions. More recently, studies have explored the use of diffusion models \cite{austin2021structured,Croitoru_Hondru_Ionescu_Shah_2022,popov2021grad} for improved data generation. For example, DiffRec~\cite{wang2023diffusion} and DiffKG \cite{diffkgwsdm} use diffusion models to denoise the interaction data and knowledge graph, respectively, leading to enhanced recommendation performance.

In contrast, our work takes a distinct approach. \model\ leverages diffusion models' denoising paradigm to refine social representations for recommendation. It encodes structural features of the social graph into low-dimensional embeddings and performs denoising with a hidden-space diffusion paradigm.

% \vspace{-0.1in}
\section{Conclusion}
\label{sec:conclusion}
This paper aims to enhance social-aware recommender systems by eliminating false or irrelevant user-wise social links. To achieve this goal, we propose \model, a novel diffusion model that trains a social denoiser through a multi-step noise propagation and elimination task. This diffusion process operates in the hidden space, utilizing encoded user representations for both effectiveness and simplicity. By training the model with varying diffusion steps, \model\ demonstrates exceptional capabilities in handling diverse noisy effects. We evaluate the effectiveness of our model through experiments on real-world datasets, showing significant improvements in recommendation accuracy compared to existing methods. In the future, we plan to explore the potential of our model in diverse recommendation scenarios, incorporating multi-modal information.
% Investigate the temporal dynamics of social relationships in the diffusion process is another avenue for future exploration.

\clearpage

\bibliographystyle{abbrv}
\balance
\bibliography{refs}

% \clearpage
% \appendix
% \input{appendix}

\end{document}